# Cell phenotypic transition proceeds through concerted reorganization of gene regulatory network

Weikang Wang[1*], Dante Poe[1,2], Ke Ni[1,2], Jianhua Xing[1,3,4,*]

[1] Department of Computational and Systems Biology, University of Pittsburgh, Pittsburgh, PA 15232, USA.

[2] Joint CMU-Pitt Ph.D. Program in Computational Biology, University of Pittsburgh, Pittsburgh, PA, USA.

[3] Department of Physics and Astronomy, University of Pittsburgh, Pittsburgh, PA 15232, USA.

[4] UPMC-Hillman Cancer Center, University of Pittsburgh, Pittsburgh, PA, USA.

- To whom correspondence should be addressed. Email: xing1@pitt.edu, weikang@pitt.edu

## Abstract

Phenotype transition takes place in many biological processes such as differentiation, and understanding how a cell reprograms its global gene expression profile is a problem of rate theories. A cell phenotype transition accompanies with switching of expression rates of clusters of genes, analogous to domain flipping in an Ising system. Here through analyzing single cell RNA sequencing data in the framework of transition path theory, we set to study how such a genome-wide expression program switching proceeds in three different cell transition processes. For each process after reconstructing a Markov transition model in the cell state space, we formed an ensemble of shortest paths connecting the initial and final cell states, reconstructed a reaction coordinate describing the transition progression, and inferred the gene regulation network (GRN) along the reaction coordinate. In all three processes we observed common pattern that the frustration of gene regulatory network (GRN), defined as overall confliction between the regulation received by genes and their expression states, first increases then decreases when approaching a new phenotype. The results support a mechanism of concerted silencing of genes that are active in the initial phenotype and activation of genes that are active in the final phenotype.

A lasting topic in science and engineering is how a dynamical system transits from one stable attractor to a new one in a corresponding state space [1]. For example, a substitution reaction in organic chemistry can proceed either through first breaking an old chemical bond to form an intermediate planar structure followed by forming a new bond (termed as a SN1 mechanism), or through forming a trigonal bipyramidal intermediate complex with breaking of the old bond and formation of the new bond take place concertedly (termed as a SN2 mechanism) [2]. Which mechanism dominates a process is determined by both the relative thermodynamic stability of the two intermediate structures, and the kinetics of forming them.

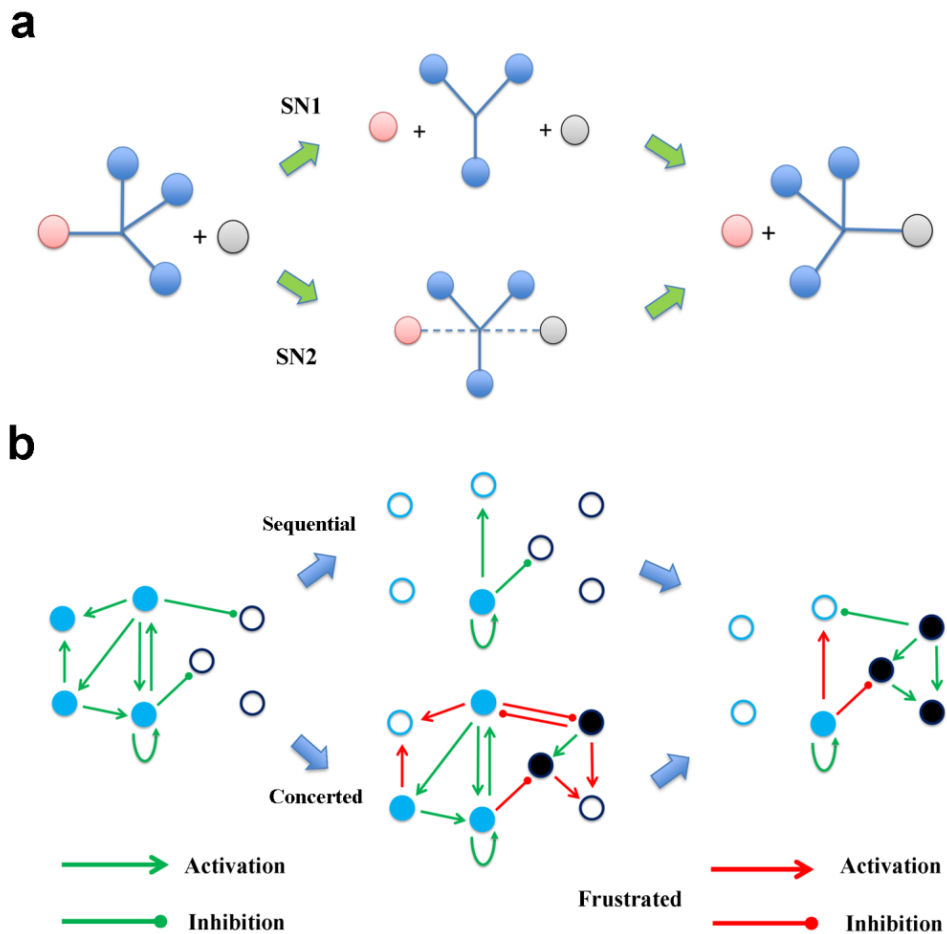

*Figure 1 State transitions may proceed through different mechanisms.* (A) SN1 and SN2 mechanisms for chemical reactions. The nodes and edges represent chemical groups and chemical bonds, respectively. (B) Sequential and concerted mechanisms for a cell phenotypic transition. Filled circles represents genes are active. Empty circle represents genes are silenced. Colors represent marker genes of different states.

Recent years there has been increasing interest in studying transitions between different cell phenotypes, partly due to available genome-wide characterization of the cell gene expression state throughout a transition process due to advances of single cell genomics techniques. A cell is a nonlinear dynamical system governed by a complex regulatory network. The latter is formed by a large number of interacting genes, and can have multiple stable attractors corresponding to different cell phenotypes. Typically a large number of phenotype-specific genes maintain a specific phenotype through mutual activation while suppressing expression of genes corresponding to other exclusive phenotypes. In some sense it resembles a spin system segregating into upward and downward domains. When a cell phenotypic transition takes place, the genes need to switch their expression status, analogous to flipping some upward and downward spin domains. A question arises as to how this transition proceeds (Fig. 1B). The transition may be sequential with gene silence first to form an intermediate with the initial cell phenotype destabilized without commitment to a new phenotype, followed by activation of other genes to instruct the cell into one specific final stable phenotype (similar to the SN1 mechanism). Alternatively gene activation and silence may happen concurrently as in the SN2 mechanism, with hybrid intermediate states co-expressing genes corresponding to the two phenotypes. One can vision two qualitatively different characteristics of the two mechanisms. Compared to the initial and final states, the intermediate states tend to have increased number of smaller subsets together with lost gene-gene interactions for the sequential mechanism, but a fewer giant clusters reflecting an increased number of gene-gene interactions for the concerted mechanism. The latter also tends to have increased number of frustrated interactions at the intermediate states. Here we borrow the concept of frustration from spin glasses, and define an interaction $F_{ij}$ from an active gene $j$ to a target gene $i$ being frustrated if the expression status of gene $i$ mismatches the nature of the interaction, e.g., activation acting on silenced gene $i$. While several existing studies have discussed frustration in undirected model network systems [3,4], here we set to examine the above two competing mechanisms against genome-wide network models reconstructed from single cell RNA-seq (scRNA-seq) datasets.

We first analyzed a scRNA-seq dataset of epithelial-mesenchymal transition (EMT) of human A549 treated with TGF-β[5], a total of $N = 3003$ single cell samples measured at several time points (Fig. S1a). During EMT cells change from epithelial to mesenchymal phenotypes with increased EMT hallmark gene set score. We selected $M = 391$ genes showing switch-like behaviors during EMT to form an $M$-dimensional state space (see SI for details). While scRNA-seq data only provide snapshots of cell transcriptomic states, a recently developed RNA velocity analysis makes it possible to extract some dynamical information [6]. RNA velocity is a high-dimensional vector that can be inferred from the quantity of spliced and un-spliced RNA, and predicts the future state of individual cells on a timescale of hours. Following Qiu et al. [7], from the RNA velocities we constructed a Markov transition model with an $N \times N$ transition matrix $T$ specifying the transition probabilities among the $N$ measured cells using a Fokker-Planck kernel (Fig.S1b).

For rate theory analyses we estimated the kernel density of the day 0 samples using Scikit-learn [8], and defined the cells whose local densities $\rho_{0d}$ are in top 100 as within the initial state A. Similarly we estimated the kernel density of the day 3 samples and selected those cells whose $\rho_{3d}$ are in top 100 as within the final state B. We randomly selected pairs of cells in state A and the state B, and obtained an ensemble of Dijkstra shortest paths between them based on the transition matrix (Fig. 2a). We applied a modified finite temperature string method [9-11] to the ensembles of transition paths (see SI for details), and obtained an array of reaction coordinate (RC) points reflecting the average dynamics of simulated shortest paths (Fig.2b). The RC points divide the *M*-dimensional state space into Voronoi cells (Fig. S2a), so the value of RC of each cell was assigned by the Voronoi cell that it locates in.

Next, to study how the regulatory network reconfigures along the RC, we need the governing dynamical equations of the EMT process. Qiu et al.[7] developed a procedure of reconstructing the generally nonlinear equations from the single cell expression vectors $\{\mathbf{x}^a\}$ and velocity vectors $\{\mathbf{v}^a = (d\mathbf{x}/dt)^a\}$, with α representing the α-th cell. Here we adopted a simpler linear model by assuming the governing equation as $\mathbf{v} = \mathbf{F}\mathbf{x} + \varepsilon$, with $\varepsilon$ being random white noises, and $F_{ij}$ quantifying the regulation of gene *j* on gene *i* so the node strength and direction in the intracellular gene regulatory network (GRN). The regulation can be direct such as gene *j* acting as a transcription factor on gene *i*, or indirect mediated through molecular species not resolved by the scRNA-seq measurements. We inferred the matrix **F**, which is in general asymmetric, from the data $\{\mathbf{x}^a, \mathbf{v}^a\}$ with the partial least square regression (PLSR) together with local false discovery rate (LFDR) methods to ensure **F** is sparse (see SI) [12].

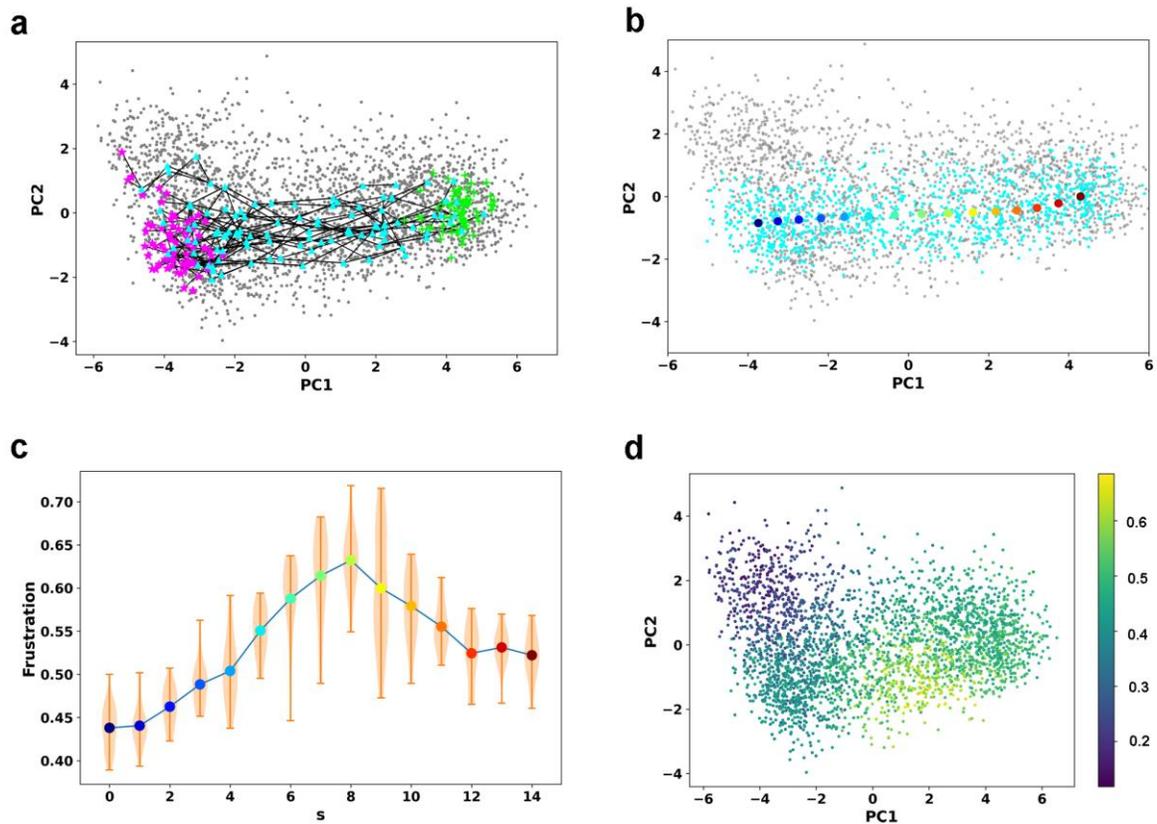

*Figure 2 Analyses on the EMT scRNA-seq data.*
*(a) Dijkstra shortest path simulation on the transition graph represented in the 2D leading PCA space. Each path is a single cell trajectory (labeled with cyan triangle) that start from the epithelial state (labeled with magenta star) and transit into the mesenchymal state (labeled with lime cross). Each dot represents a cell.*
*(b) RCs calculated from the Dijkstra shortest paths. The cyan dots are cells close to the RC to form a reaction tube, i.e., cells within each Voronoi cell that are k-nearest-neighbors of the corresponding RC point.. These cells are used to infer the **F** matrix.*
*(c) Frustration score along the RC (s) of EMT.*
*(d) Cell-specific variation of effective regulation edges in the GRN of EMT. Colors represent proportions of effective regulation edges to all edges in the GRN within each cell (represented as a dot).*

With **F** being the same for all cells, the effective GRN may differ between different cells. Notice that a prerequisite for gene *j* acting on *i* is that gene *j* is expressed in the cell, otherwise the *j* → *i* edge is treated as non-exist in this specific cell. Therefore with the gene expression binarized as 0 for silence, and 1 for active expression, we defined a frustration score for the interaction between a pair of genes (i, j) as $fs_{ij} = s_j \text{sgn}((2s_i - 1)F_{ij})$, assuming a value 1 (not frustrated), 0 (no regulation), and -1 (frustrated), and sgn(x) is the usual sign function,

$$\text{sgn}(x) = \begin{cases} 1 & \text{if } x > 0 \\ 0 & \text{if } x = 0 \\ -1 & \text{if } x < 0 \end{cases}.$$

Furthermore we defined the overall frustration score of a cell-specific GRN as the fraction of frustrated edges out of all nonzero edges (i.e., $fs_{ij} \neq 0$) in the whole network of the cell. For EMT, the average frustration score along RC increases first and reaches the peak when cells were treated with TGF-β for about one day, then decreases (Fig. 2c), consistent with the concerted but not the sequential mechanism. We also calculated $H = -\sum_{i,j} fs_{ij}$, analogous to the pseudo-Hamiltonian of a cell defined by Font-Clos et al [3] but with directed regulation, which also shows similar peaked profile along the RC (Fig. S2b).

To further evaluate the two possible mechanisms, we calculated the number of effective edges, i.e., edges with nonzero $F_{ij}$ and gene j expressed ($s_j = 1$) in the cell. The number of effective edges increases first and then decreases during EMT (Fig. 2d). This initial increase of effective edges reflect transient co-activation of marker genes of state A and B and existence of their cross-regulation (Fig. S2c), and the active edge number then decreases with silence of epithelial genes.

To rule out the possibility that the observed properties are specific for the Markov model we used, we repeated the above analyses with a transition matrix reconstructed from a correlation kernel of Bergen et al.[13] The analyses gave similar RC, and the frustration score as well as the number of effective edges (Fig. S3) change along the RC similarly as observed with the model obtained using that of Qiu et al.

To investigate whether the concerted mechanism is general for CPTs, we performed the same analyses on two additional CPT scRNA-seq datasets. One is on development of pancreatic endocrine cells [14]. During embryonic development Ngn3-low progenitors first transform into Ngn3-high precursors then Fev-high cells. The latter further develop into endocrine cells, specifically glucagon producing α-cells that we focus on here (Fig. S4). The cells in state A and B were selected from Ngn3-low progenitors and glucagon producing α-cells, respectively. Along the RC (Fig. S5a), the peak of average frustration score locates at the cell population with high Ngn3 expression (Fig. 3a Left). The pseudo-Hamiltonian value shows trend similar to the frustration score (Fig. S2b). The number of effective edges increases first and then decreases while low Ngn3 expression cells transit into α-cells (Fig. 3a Right). The number of interactions between marker genes of Ngn3-low progenitors and glucagon producing α-cells show similar trend with the active edges (Fig. S5c).

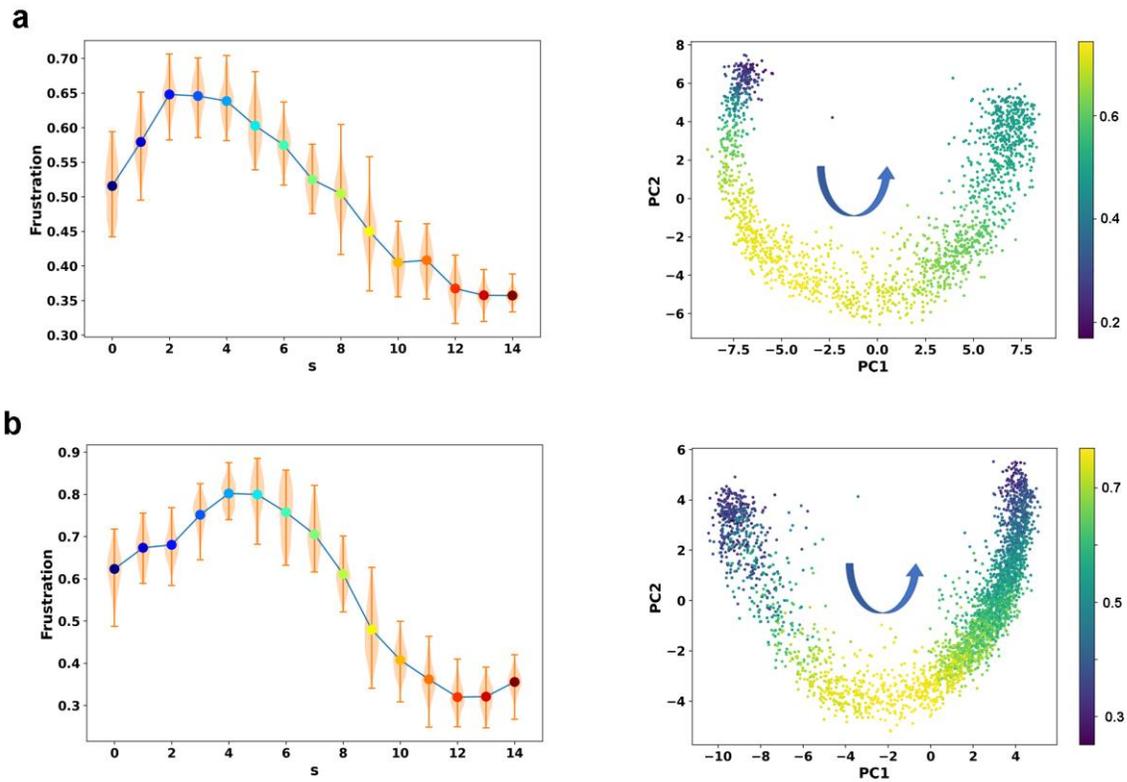

*Figure 3 Analyses on pancreatic endocrinogenesis and dentate gyrus neurogenesis.*
*(a) Frustration score along the RC (left) and cell-specific variation of effective regulation edges in endocrine cell development (right). Colors represent proportions of effective regulation edges within each cell to all edges in the GRN. Arrow represents the direction of development.*
*(b) Same as in panel (a) except for the granule cell lineage development dataset.*

Another system is on development of granule cell lineage in dentate gyrus, where radial glia-like cells differentiate through nIPCs, Neuroblast 1 and 2, immature granule cells, and eventually into mature granule cells (Fig. S6) [15]. Along the calculated RC (Fig. S7a), the neuroblast has highest frustration score (Fig. 3b left). The pseudo-Hamiltonian value exhibits dynamics similar to the frustration score (Fig. S7b). The number of effective edges increase first and then decrease when low radial glia like cells transit into mature granule cells (Fig. 3b Right). The number of interactions between marker genes of radial glia-like cells and mature granule cells show similar trend with the active edges (Fig. S7c).

Therefore the dynamical properties of the two processes are consistent with the concerted mechanism. Compared with the EMT results, in these two developmental processes the frustration score of state A is higher than that of state B. As the Ngn3-low progenitors and radial glia-like are stem-like cells, this observation suggests that the GRN of stem-like cells has higher frustration than that of differentiated cells, consistent with a widely assumed hypothesis that stem cells are close to a critical point.

The idea of relating CPTs and chemical reactions has been discussed in the literature [16]. Here we presented a procedure of reconstructing the RC of a CPT process from scRNA-seq data. A related concept is the transition state. In chemical reactions it typically refers to short-lived intermediates or a state of maximal potential energy along the RC. It is tempting to identify the intermediate state with highest frustration as the "transition state", while it is unclear whether it is indeed a dynamical bottleneck of the associated CPT process.

In summary, in this work through analyzing scRNA-seq data of CPTs in the context of dynamical systems theory we identify that many CPTs may share a common concerted mechanism. This conclusion is also supported by an increasing number of studies on various CPT processes reporting existence of intermediate hybrid phenotypes that have co-expression of marker genes of both the initial and final phenotypes such as the partial EMT state [17]. Notice that a cell typically has multiple target phenotypes to choose, functionally the concerted mechanism may allow canalized transition for directing the cells to transit to a specific target phenotype, as visioned by the developmentalist C. H. Waddington [18].

**Supplemental Information**

**1 Gene selection**

In this paper, we focus on those genes showing switch-like behavior during the phenotype transition. First we select the high-expression gene across the whole dataset. To avoid bias selection, we also select the high-expression genes in each cell type. In EMT dataset, we select the high-expression gene in 0d, 8h,1d and 3d separately. The filtering criteria is minimum number of counts and minimum number of cells. The minimum number of counts is 20. And minimum number of cells is set as 5% of the number of cells of the corresponding cell type. And top 2000 high-expression genes of the whole dataset are kept. The gene set is the union set of these selected genes.

For the pancreatic endocrinogenesis, the minimum number of counts is 20 and minimum number of cells is set as 10% of the number of cells of the corresponding cell type. Top 1000 high-expression genes of the whole dataset are kept. The union set of these genes is used for later calculation.

For the Dentate gyrus neurogenesis, the minimum number of counts is 20 and minimum number of cells is set as 10% of the number of cells of the corresponding cell type. Top 1000 high-expression genes of the whole dataset are kept for subsequent analyses. The union set of these genes is used for later calculation.

We compare the gene distributions in state A and state B. For EMT, the comparison is between the sample at day 0 and the sample treated with TGF-β for three days. For the pancreatic endocrinogenesis, the comparison is between Ngn3-low progenitors and glucagon producing α-cells. For the Dentate gyrus neurogenesis, the comparison is between radial glia-like cells and mature granule cells. If the distribution shows significant shift, i.e., the distance between peaks of distributions is larger than sum of half width of both peaks between any two time points, $D_{t_1,t_2} > \left(h_{w,t_1} + h_{w,t_2}\right)/2$), this gene is selected. The threshold is set as the weighted mean of the two distributions. At each time point, the genes are selected based on minimum shared counts and minimum shared cells. The genes that can be binarized are selected from the union set of the genes at all time points.

**2 path analysis from single cell RNA velocity analysis**

With scRNA-seq velocity analysis, we reconstructe the velocity graph of the whole cell population, which is a transition matrix between all pairs of the cells. Each cell is treated as a node in this network. The distance between different cells is $-log\ P_{ij}$, where $P_{ij}$ is the transition probability between cells. Here we add one constrain in the velocity graph that only the transition between the cells that their sample time points are successive or they have the same sample time. We also take the cell density at each sample time point into consideration. The density of each single cell $\rho_t$ (normalized by the maimum) at its sample time is evaluated by kernel density estimation. The transition from cell $i$ to cell $j$ is penalized by the relative cell density of cell $j$. The distance from cell $i$ to cell $j$ with penalty is $-log\, P_{ij} \times (1 + 5 * \exp(-(\rho_t - r)))$. For EMT, r is 0.5. 0.8 is used for the other two datasets. The high-density region of each sample time is more likely to be passed by the single cell trajectories. 100 cell pairs $(c_f, c_l)$ are randomly selected from high density regions in the first and last sample populations. The shortest paths between the cell pairs are calculated with Dijkstra's algorithm. These shortest paths are the probable single cell trajectories in the phenotype transition.

**3 Procedure for determining a RC**

We follow a procedure adapted from what used in the finite temperature string method for numerical searching of RC and non-equilibrium umbrella sampling [9,19].

a) Identify the starting and ending points of the reaction path as the means of data points in the state A and state B, respectively. The two points are fixed in the remaining iterations.
b) Construct an initial guess of the reaction path that connects the two ending points in the feature space through linear interpolation. Discretize the path with $N$ (= 15) points (called images, and the $k_{th}$ image denoted as $s_k$ with corresponding coordinate $\mathbf{X}(s_k)$) uniformly spaced in arc length.
c) For a given trial RC, divide the multi-dimensional state space by a set of Voronoi polyhedra containing individual images, and calculate the score function
$F = \sum_k \sum_u \sum_\alpha \sqrt{\|\mathbf{X}(s_k) - \mathbf{X}_{u,t_\alpha} \mid \mathbf{X}_{u,t_\alpha} \in s_k\|^2} + w \sum_k \sum_u d_{k,u}$. where $\mathbf{X}_{u,t_\alpha}$ stands for the points on simulated trajectory $u$ at step $t_\alpha$ that reside within the $k_{th}$ polyhedron (containing image point $s_k$); $d_{k,u}$ is the distance between image $s_k$ and trajectory $u$, defined as the distance between each image on the path to the closest point on the trajectory, $d_{k,u}^2 = \arg\min \|\mathbf{X}(s_k) - \mathbf{X}_{u,t_\alpha}\|^2$; $w$ is a parameter that specifies the relative weights between the two terms in the right hand of the expression, here we use 2.
d) We carry out the minimization procedure through an iterative process. For a given trial path defined by the set of image points, we calculate a set of average points using the following equations, $\overline{\mathbf{X}}(s_k) = \dfrac{\left\langle \sum_u \sum_\alpha \{\mathbf{X}_{u,t_\alpha} \mid \mathbf{X}_{u,t_\alpha} \in s_k\} \right\rangle + w \left\langle \sum_u \mathbf{X}_{u,\arg\min\|\mathbf{X}(s_k) - \mathbf{X}_{u,t_\alpha}\|^2} \right\rangle}{1 + w}$. Next we update the continuous reaction path through cubic spline interpolation of the average positions [20], and generated a new set of $N$ images $\{X(s_k)\}$ that are uniformly distributed along the new reaction path. We set a smooth factor, *i.e.*, the upper limit of $\sum_{k=1}^{N}(\overline{X}(s_k) - X(s_k))$, as 1 for calculating the RC.
e) We iterate the whole process in step 3 until there was no further change of Voronoi polyhedron assignments of the data points.

# 4 Network inference

We adopt a partial linear square regression (PLSR) method to infer the gene regulation networks. The velocity vector of each single cell $d\boldsymbol{x}/dt$ and the level of spliced mRNA $\boldsymbol{x}$ are used for inferring the GRN with PLSR method. $d\boldsymbol{x}/dt = \boldsymbol{Fx} + error$, where $\boldsymbol{F}$ is constant and generally asymmetric matrix describing gene regulation strength. Regression methods are widely used in network inference, and among which the PLSR has several advantages. First, it can be used when the number of features is larger than the number of samples. In the scRNA dataset, the number of genes is often comparable to or larger than the number of cells. Second, it can avoid over-fitting because it uses major components for regression. However, the regulation relation $F$ obtained from PLSR is typically a dense matrix, while most GRNs are sparse. To generate a sparse network, we further adopt the method of local false discover rate (LFDR) to select those regulation relations that are statistically significant. This procedure ensures the GRN is sparse [12]. Since cells within each Voronoi cell can scatter dispersively in the orthogonal space, we select only cells close to the RC for infering the $\boldsymbol{F}$ matrix. That is, among cells within each Voronoi cells, we select the k-nearest-neighboring (KNN) cells of the corresponding RC point. Such cells from all Voronoi cells collectively form the set for $\boldsymbol{F}$ matrix inference. The inference is performed using scikit-learn [8] by maximizing the covariance between $\boldsymbol{x}$ and $d\boldsymbol{x}/dt$ in the PLSR method. The value of components is 2 and data are scaled. In LFDR, the null hypothesis $H_0$ assumes that $F_{i,j}$, which is regulation from gene $j$ to gene $i$, is 0. An interaction is identified as nonzero when $fdr(F_{i,j}) < q$, where $fdr(F_{i,j})$ is the false discover rate and $q$ is the threshold [21]. The following R package is used for calculation (*https://rdrr.io/cran/locfdr/*). The central matching estimation method is used. The degrees of freedom is set as 10 in calculation, and q is set as 0.1.

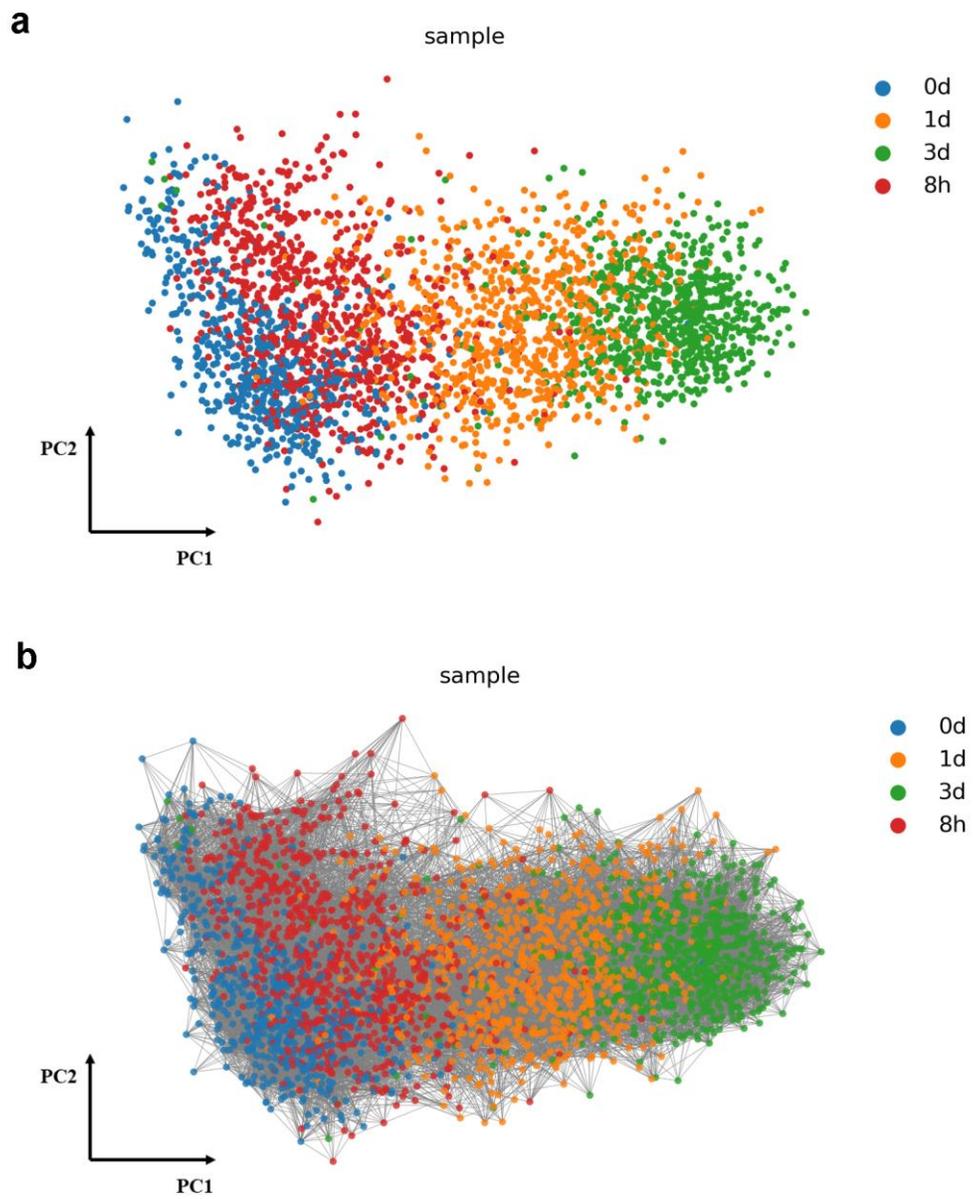

***Figure S1 scRNA-seq datasets of EMT in reduced dimension.***
*(a) Scatter plot in the 2D leading PCA space.*
*(b) Transition graph based on RNA velocity.*

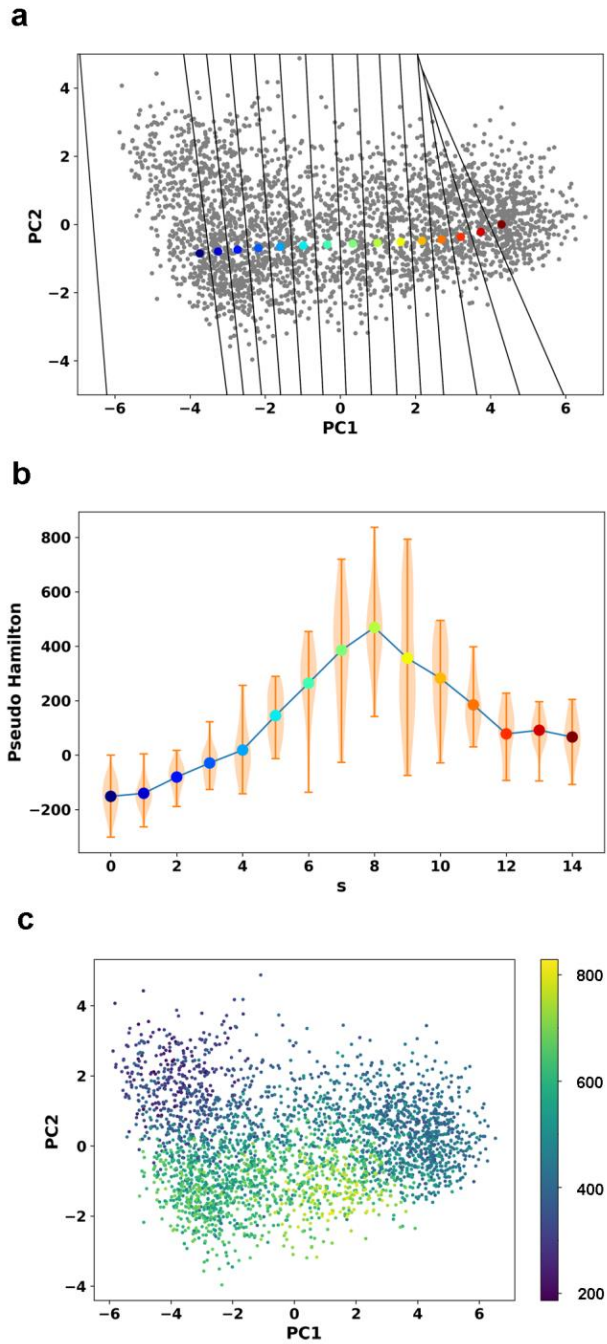

***Figure S2 Additional results on transition path analyses of the EMT dataset.***
*(a)Voronoi cells defined by an array of points equally distributed along the RC divide the feature space into different regions.*
*(b) Pseudo Hamilton values along the RCs.*

*(c) Cell-specific number of interactions between maker genes of epithelial state and mesenchymal state of all cells.*

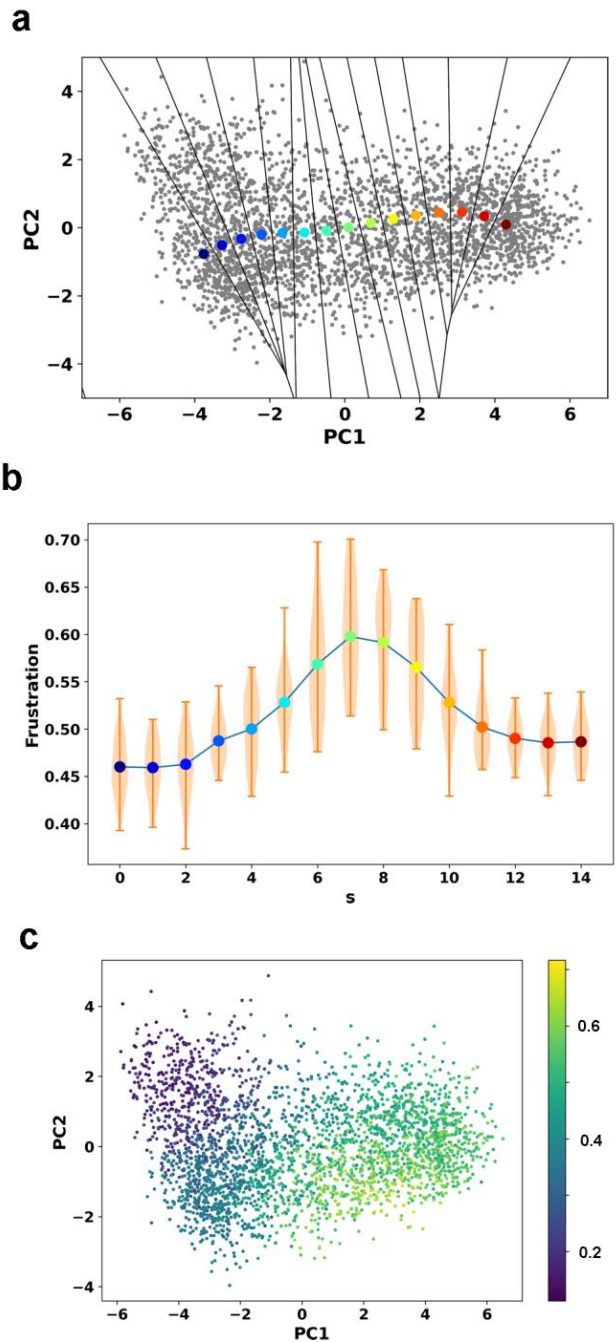

***Figure S3 Analysis of the EMT dataset with the package scvelo.***
*(a) The RC calculated from the Dijkstra shortest paths.*
*(b) Frustration score along the RC of EMT.*

*(c) Variation of effective regulation edges in the GRN in the processes of EMT. Colors represent proportions of effective regulation edges in individual cells relative to all edges in the GRN.*

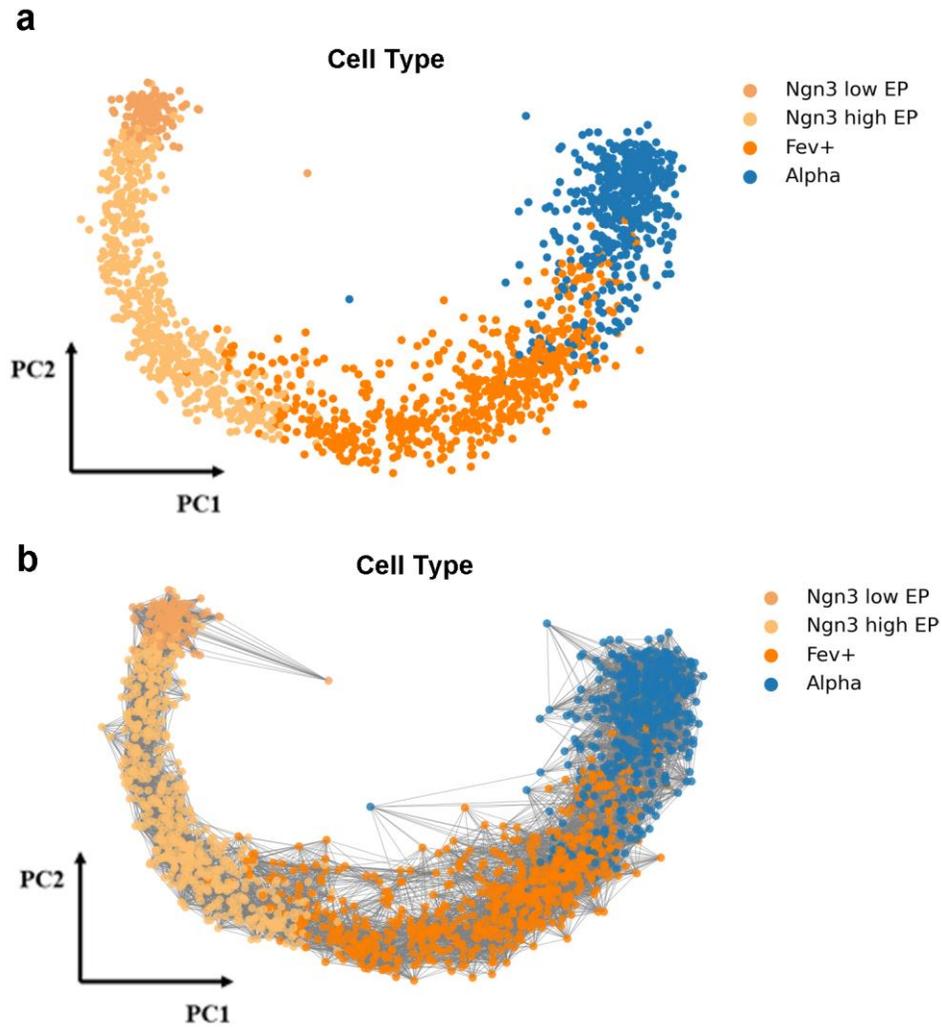

***Figure S4 scRNA-seq datasets of pancreatic endocrinogenesis. The number of cells is 1731. 681 genes are selected.***
*(a) Scatter plot in the 2D leading PCA space.*
*(b) Transition graph based on RNA velocity.*

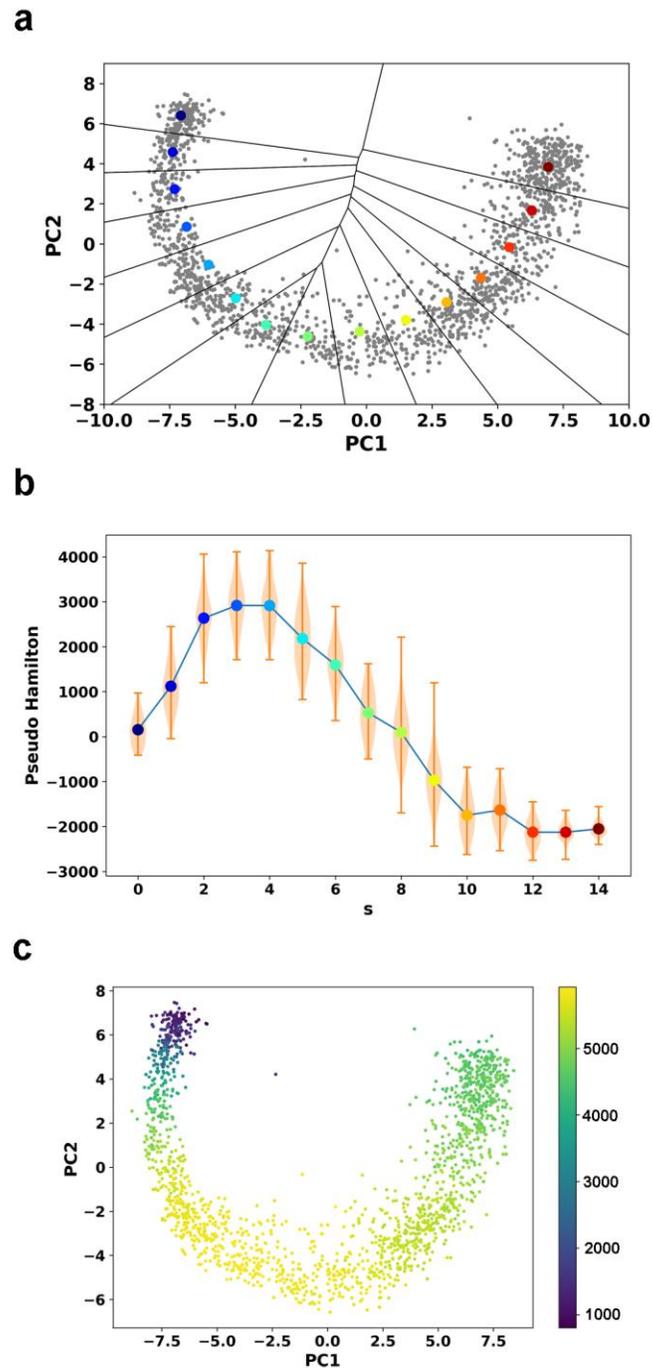

***Figure S5 Transition path analyses of the pancreatic development dataset.***
*(a) Voronoi cells defined by an array of points equally distributed along the RC divide the expression space of pancreatic endocrinogenesis into different regions.*
*(b) Pesudo-Hamiltonian values along the RC.*

*(c) Cell-specific number of interactive edges between marker genes of Ngn3-low progenitors and glucagon producing α-cells in all cells. Each dot represents a cell.*

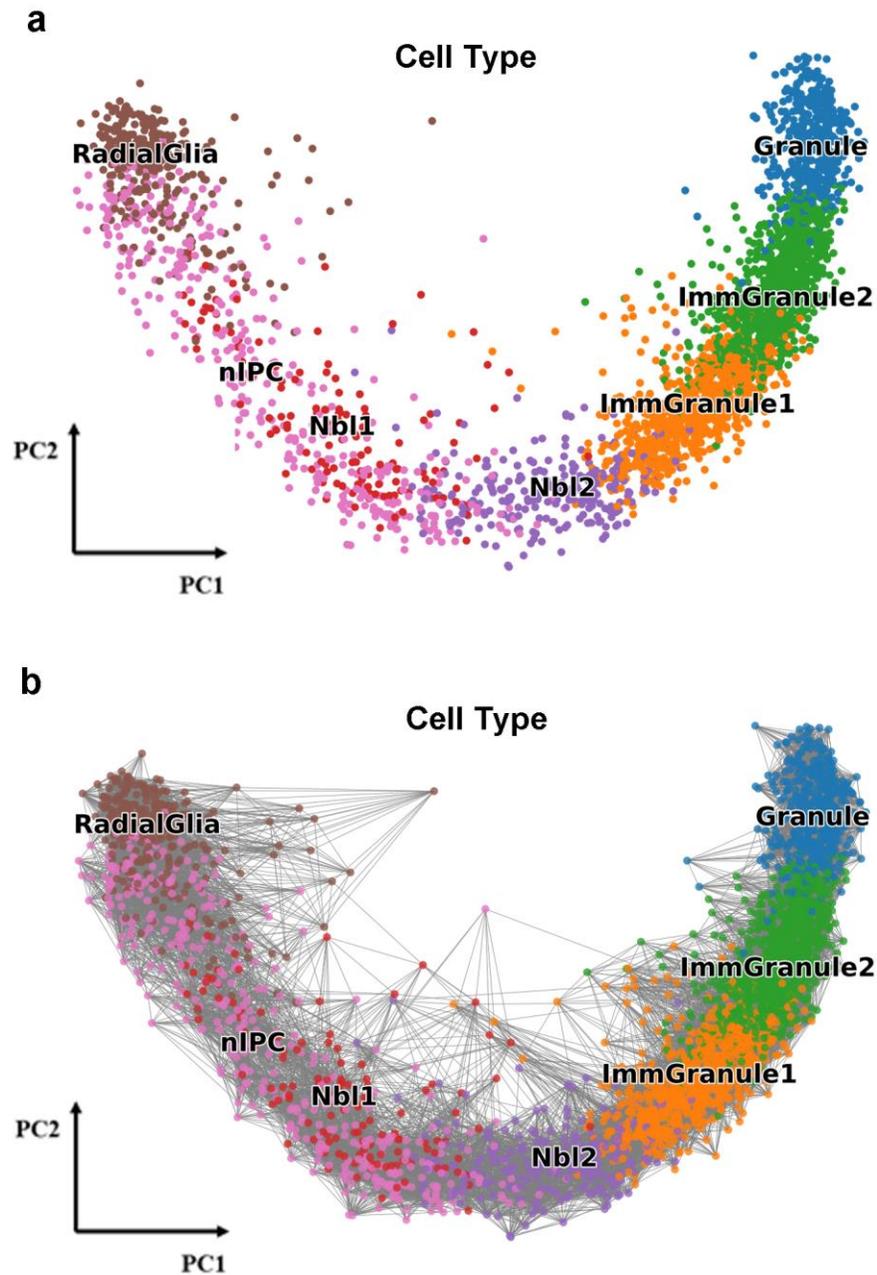

***Figure S6 scRNA-seq dataset of dentate gyrus neurogenesis. The age of this sample is P5. The number of cells is 3179, and 1078 genes are selected.***
*(a) Scatter plot in the 2D leading PCA space.*
*(b) Transition graph based on RNA velocity.*

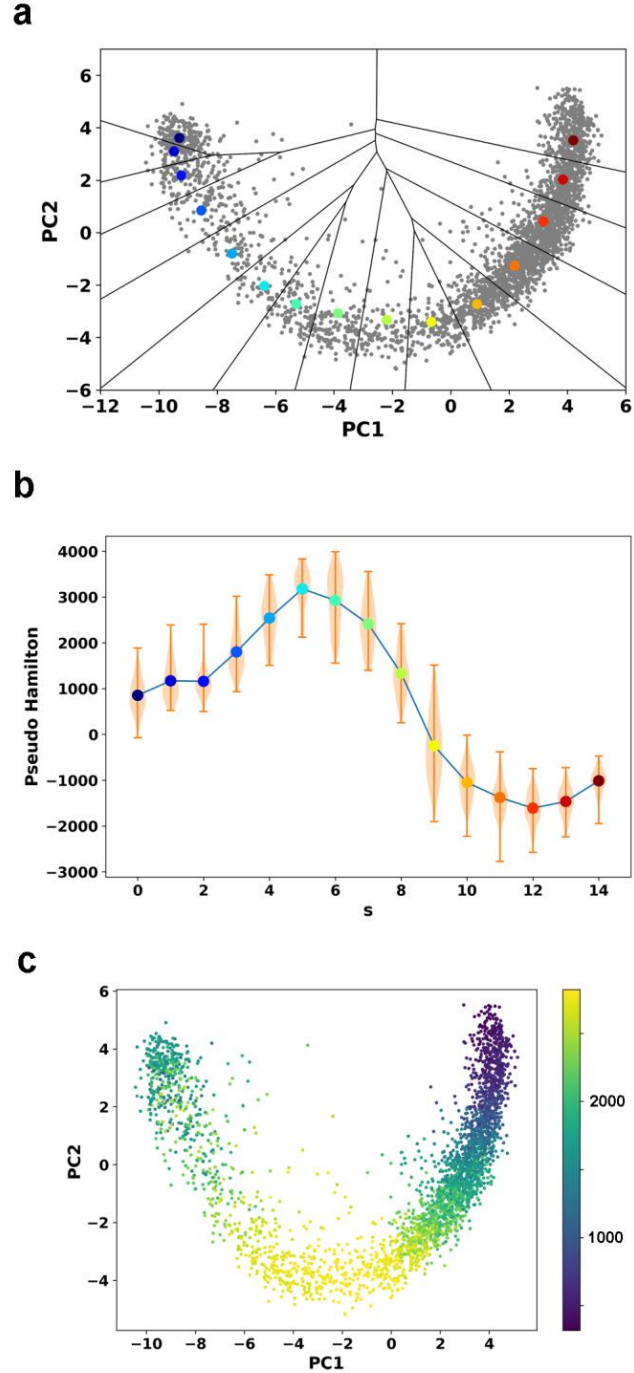

***Figure S7 Transition path analyses of the dentate gyrus neurogenesis dataset.***
*(a) Voronoi cells defined by an array of points equally distributed along the RC divide the expression space of dentate gyrus neurogenesis into different regions.*
*(b) Pesudo-Hamiltonian values along the RC.*

*(c) Cell-specific number of interactive edges between marker genes of radial glia-like cells and mature granule cells in all cells. Each dot represents a cell.*